# Independent tuning of electronic properties and induced ferromagnetism in topological insulators with heterostructure approach

Zilong Jiang[†], Cui-Zu Chang[‡], Chi Tang[†], Peng Wei[‡], Jagadeesh S. Moodera[‡, §], and Jing Shi[†*]

[†]Department of Physics and Astronomy, University of California, Riverside, CA 92521

[‡]Francis Bitter Magnetic Lab, Massachusetts Institute of Technology, Cambridge, MA 02139

[§]Department of Physics, Massachusetts Institute of Technology, Cambridge, MA 02139

**ABSTRACT:** The quantum anomalous Hall effect (QAHE) has been recently demonstrated in Cr- and V-doped three-dimensional topological insulators (TIs) at temperatures below 100 mK. In those materials, the spins of unfilled *d*-electrons in the transition metal dopants are exchange coupled to develop a long-range ferromagnetic order, which is essential for realizing QAHE. However, the addition of random dopants does not only introduce excess charge carriers that require readjusting the Bi/Sb ratio, but also unavoidably introduces paramagnetic spins that can adversely affect the chiral edge transport in QAHE. In this work, we show a heterostructure approach to independently tune the electronic and magnetic properties of the topological surface states in $(Bi_xSb_{1-x})_2Te_3$ without resorting to random doping of transition metal elements. In heterostructures consisting of a thin $(Bi_xSb_{1-x})_2Te_3$ TI film and yttrium iron garnet (YIG), a high Curie temperature (~ 550 K) magnetic insulator, we find that the TI surface in contact with YIG becomes ferromagnetic via proximity coupling which is revealed by the anomalous Hall effect (AHE). The Curie temperature of the magnetized TI surface ranges from 20 to 150 K but is uncorrelated with the Bi fraction *x* in $(Bi_xSb_{1-x})_2Te_3$. In contrast, as *x* is varied, the AHE resistivity scales with the longitudinal resistivity. In this approach, we decouple the electronic properties from the induced ferromagnetism in TI. The independent optimization provides a pathway for realizing QAHE at higher temperatures, which is important for novel spintronic device applications.

**TOC GRAPHIC** 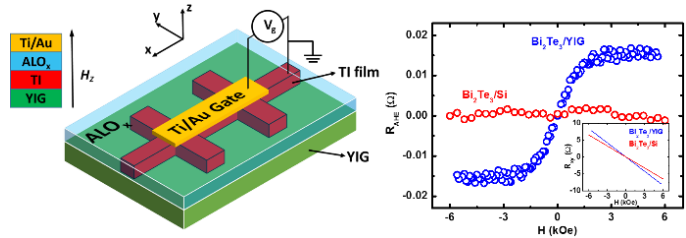

Quantum anomalous Hall effect (QAHE) requires both a spontaneous ferromagnetic order and a topological nontrivial inverted band structure.[1-3] To introduce the ferromagnetic order, random doping of transition metal elements, e.g. Cr or V, has been employed.[4-7] Although the Curie temperature ($T_C$) of the magnetic TI can be as high as ~ 30 K,[4,7] QAHE only occurs at temperatures two orders of magnitude below $T_C$.[7-10] While the mechanism of this large discrepancy remains elusive, in order to observe QAHE at higher temperatures, it is essential that the exchange interaction in magnetic TI is drastically increased; in the meantime, the magnetic disorder needs to be greatly reduced. For the random doping approach, it is rather difficult to accomplish these two objectives. An alternative way to address both issues simultaneously is to couple a non-magnetic TI to a high $T_C$ magnetic insulator to induce strong exchange interaction via the proximity effect.[11-13]

The magnetic proximity effect is a well-known phenomenon that has been intensely investigated.[14-20] By proximity coupling, the surface layer of TI acquires a magnetic order without being exposed to any random magnetic impurities.[16-18] This heterostructure approach was previously adopted in ref. [16] using EuS, a ferromagnetic insulator with a band gap of 1.6 eV and a $T_C$ of ~ 16 K, grown on a 20 quintuple-layer (QL) thick $Bi_2Se_3$. Both magnetoresistance and a small AHE signal at low temperatures were ascribed to an induced magnetization in $Bi_2Se_3$. Alternatively, YIG has a much larger band gap (~ 2.85 eV) and a much higher $T_C$ (~ 550K), and therefore is a better magnetic insulator for heterostructures. In $Bi_2Se_3$/YIG, the magnetoresistance observed at low temperatures indicates an interaction effect between the two materials.[17,18] We observed suppressed weak anti-localization in 20 QL-$Bi_2Se_3$ on YIG.[17] Based on our recent successes in Pd/YIG,[20] $Bi_2Se_3$/YIG,[17] and graphene/YIG,[15] here we demonstrate that high-quality heterostructures of 5 QL thick $(Bi_xSb_{1-x})_2Te_3$ grown on atomically flat YIG films exhibits induced ferromagnetism at the TI surface. By varying the ratio of Bi to Sb, we can effectively tune the electronic properties such as the carrier density and resistance of the TI layer without affecting the magnetic properties of YIG or the induced magnetic layer in TI.

The atomically flat YIG films (~20 nm) were first epitaxially grown on (111)- gadolinium gallium garnet (GGG) substrates via pulsed laser deposition as described previously.[17,20] Room temperature ferromagnetic resonance (FMR), vibrating sample magnetometry (VSM), and atomic force microscopy (AFM) measurements have been performed on all samples (see Supporting Information). YIG films show clear in-plane magnetic anisotropy (Fig. 1b) and the bulk magnetization value ($4\pi M_s$ ~ 2000 Oe). Then they were transferred to an ultra-high vacuum molecular beam epitaxy (MBE) system for TI film growth. After high temperature annealing (degasing), 5 QL-thick TI films were grown on YIG and then capped with a 10 nm thick epitaxial Te layer. The sharp and streaky reflection high energy electron diffraction (RHEED) pattern (Fig. 1d) of the 5 QL TI indicates a flat surface and well-defined single crystal structure. High single crystal quality was also confirmed by x-ray diffraction (XRD) on a 20 QL TI on (111)-oriented YIG/GGG, as shown in Fig. 1e. All peaks can be identified with (00n) diffraction of $(Bi_xSb_{1-x})_2Te_3$, while the YIG/GGG shows the (444) diffraction peak and the (001) peak of the Te capping layer is also present. No other phase is observed in the TI/YIG film according to the XRD data. The zoom-in low-angle XRD scan near the (003)-peak shows multiple Kiessig fringes on both sides, further revealing excellent layered structures of TI films on (111)-oriented YIG and good TI/YIG interface correlation. For transport studies, Hall bars of 900 μm ×100 μm were fabricated by photolithography and etching the TI layers by inductively coupled plasma (ICP). For selected samples, a 50 nm thick $Al_2O_3$ layer was grown as a top gate dielectric by atomic layer deposition (ALD), and a 80 nm thick Ti/Au layer was deposited by electron beam evaporation to form a top-gated device (Fig. 1a).

In this work, we have prepared multiple $(Bi_xSb_{1-x})_2Te_3$ samples with six different Bi fractions, *i.e.* $x$=0, 0.16, 0.24, 0.26, 0.36, and 1. By varying Bi content, the carrier concentrations of TI samples are systematically controlled,[21,22] so that the position of the Fermi level is tuned from the bulk valence band (e.g. $x$=0), through the band gap (intermediate $x$'s), and to the bulk conduction band (e.g. $x$=1). The Fermi level position tuning allows us to control the relative contributions to electrical transport from the bulk and surface states. For the surface state-dominated samples (e.g. $x$=0.24), we can continuously fine tune the Fermi level across the Dirac point by electrostatic gating.

Figure 2a displays the temperature dependence of five TI/YIG samples. For the $Bi_2Te_3$ ($x$=1) and $Sb_2Te_3$ ($x$=0) samples at the extreme doping levels, the resistivity ($R_{xx}$) is lower than that of the other three samples over the entire temperature range. Moreover, these two samples show metallic behaviors, *i.e.* $dR_{xx}/dT > 0$ over the most temperature range, while the other three samples show a stronger insulating tendency, *i.e.* $dR_{xx}/dT < 0$, due to depletion of bulk carriers at lower temperatures. Among these three insulating samples, the $x$=0.26 sample has the highest $R_{xx}$ at 2 K, reaching 8.4 kΩ. Fig. 2b summarizes both the 2 K resistivity and the carrier density vs. the Bi fraction $x$. As $x$ increases from 0 to ~ 0.16, $R_{xx}$ increases and the carriers are holes from the Hall measurements. The hole carrier density decreases as $x$ approaches 0.16. The increasing $R_{xx}$, decreasing 2D carrier density, and the insulating behavior of $x$= 0.16 sample all indicate that the Fermi level shifts up from the bulk valence band into the bulk band gap.[21] As $x$ increases further, the $R_{xx}$ continually increases, and the 2D hole density passes the minimum and then carriers switch to electrons. These facts suggest that the Fermi level passes the Dirac point of the topological surfaces states between $x$~ 0.16 and 0.26. As $x$ increases further, the Fermi level shifts up more and finally enters the bulk conduction band as $x$ approaches 1. Fig. 2c illustrates a schematic band diagram when $x$ is varied.[23] The actual band structure and the precise Fermi level position for each $x$ require detailed first-principles calculations. Nevertheless, the relative position of the Fermi level with respect to the Dirac point for different $x$ values can be qualitatively determined from our experimental data.

To probe the proximity induced ferromagnetism in the TI surface contacting YIG, we focus on the nonlinear Hall signal by removing the dominant linear ordinary Hall background signal.[15,16] The inset of Fig. 3a shows a comparison between $Bi_2Te_3$/YIG and $Bi_2Te_3$/Si. Both have strong linear Hall signals with negative slopes as expected for an n-type $Bi_2Te_3$. The carrier density of TI layer is $4.4 \times 10^{13}$/cm$^2$ for $Bi_2Te_3$/YIG and $5.4 \times 10^{13}$/cm$^2$ for $Bi_2Te_3$/Si, which is not very sensitive to substrate. However, as the linear background is removed, the remaining Hall signal from the two samples shows a distinct difference. While $Bi_2Te_3$/Si does not have any definitive nonlinear signal left, $Bi_2Te_3$/YIG has a clear nonlinear component with a saturation feature. In general, non-linearity in Hall voltages can arise from co-existing two types of carriers. In fact, such non-linearity is often observed when the Fermi level is in the vicinity of the Dirac point where both electrons and holes are present.[24] These two samples are clearly in the single carrier type regime; therefore, we exclude the two-carrier possibility. The shape of the nonlinear Hall signal in $Bi_2Te_3$/YIG resembles that of the YIG hysteresis loop in perpendicular magnetic fields (Fig. 1b). We thus assign this nonlinear signal a contribution from AHE. Further evidence will be discussed shortly. Although it is not straightforward to determine the exact physical origin of the observed AHE, it is known that AHE must stem from ferromagnetism in conductors.[25] Since the underlying YIG is found to remain insulating (resistance >40 GΩ) when measured after the TI growth, etching, and device fabrication are completed, we exclude that the YIG surface itself becomes conducting and contributes to the AHE signal. Moreover, since YIG is grown at ~700 ºC while TI is grown at ~ 250 ºC later, we do not expect any significant diffusion of Fe atoms to dope the TI to turn it to ferromagnetic. Hence, we conclude that the bottom metallic surface of the TI film becomes ferromagnetic via the proximity coupling just as what has been observed in other systems.[15,16,19,20]

Note that in the $Bi_2Te_3$ sample ($x=1$) the saturation value of the AHE magnitude $R_{AHE}$ is only ~ 0.015 Ω, much smaller than $R_{xx}$. In fact, all samples show clear AHE signals at 2 K. More importantly, as the carrier type switches as $x$ goes from 0.16 to 0.26, the sign of the AHE resistivity remains the same. Since the ordinary Hall effect arises from the Lorentz force associated with an external magnetic field such as the stray magnetic field from domain boundaries, if the nonlinear Hall signal observed here is due to the ordinary Hall effect, its sign would change as the carrier type switches. The absence of the sign change further confirms the AHE nature of the nonlinear Hall signal, *i.e.* it is a consequence of the induced ferromagnetic surface of TI.[25] Moreover, $R_{AHE}$ follows the same trend as that of $R_{xx}$ (Fig. 3b), i.e. the more insulating samples showing larger $R_{AHE}$. In $x= 0.26$ sample, $R_{AHE}$ jumps to nearly 2 Ω, over two orders of magnitude larger than in $Bi_2Te_3$ (*x*=1). After passing the crossover point, $R_{AHE}$ decreases to 1.5 Ω at *x*=0.16 and finally drops to 0.12 Ω at *x*=0 ($Sb_2Te_3$) which is only 10% of the maximum value. Quantitatively, the correlation between $R_{xx}$ and $R_{AHE}$ can be better seen in Fig. 3c where a power-law with an exponent ~ 2 best fits the data. Since $\rho_{xy} \ll \rho_{xx}$, and $\sigma_{xy} = -\frac{\rho_{xy}}{\rho_{xx}^2 + \rho_{xy}^2}$, this power-law suggests that the AHE conductivity is nearly independent of $\rho_{xx}$. Since *x* is the controlling parameter, the quadratic relation between $\rho_{xx}$ and $\rho_{xy}$ suggests that $\sigma_{xy}$ is constant (shown in the inset of Fig. 3c), which rules out the skew scattering mechanism.

The induced ferromagnetism arises from the hybridization between the boundary layers of the two materials in the TI/YIG heterostructures; therefore, the resulting exchange coupling is expected to be weaker than that in the interior of YIG.[16] Additionally, less than ideal interfaces can further weaken the exchange coupling strength. To quantify the exchange interaction of the proximity effect, we measure the AHE signal as the temperature is increased until it vanishes. We define it as the ferromagnetic ordering temperature or $T_C$ for the induced magnetic layer in TI. The $T_C$ of all TI/YIG samples is above 20 K and can be as high as ~ 150 K, as shown in Fig. 3d. There seems to be no correlation between $T_C$ and *x* or the carrier concentration (see Fig. S5). Instead, this sample-to-sample variation may be attributed to variations in the state of the TI-YIG interface. Although it is not possible to pinpoint the most important factor (e.g. oxidation state and surface termination) responsible for the exchange strength, there is no fundamental reason that the $T_C$ should be limited to 150 K. Future improvement of interface quality is expected to result in a higher $T_C$ of the magnetized layer.

For the insulating samples whose Fermi level is located in the bulk band gap of the TI, we can further fine tune the position with a gate to access the surface states at different energies. Fig.

4a shows the gate voltage dependence of $R_{xx}$ for the $x=0.24$ sample which has a top gate above a 50 nm thick $Al_2O_3$ insulator. At zero gate voltage, *i.e.* $V_g=$ 0 V, the temperature dependence shows a bulk insulating behavior and the ordinary Hall data indicates the *p*-type conduction. Therefore, the Fermi level is located just below the Dirac point in the band gap. As the gate voltage is swept from negative to positive, the resistivity reaches a peak ($R_{xx}$ ~ 22 kΩ) at $V_g$ ~ 25 V and the carrier type switches from the *p*- to *n*-type as indicated by the ordinary Hall background. As $V_g$ is swept, the Fermi level moves upwards and passes the Dirac point which coincides with the maximum in resistivity. The AHE data for two representative gate voltages, *i.e.* 0 and 40 V, are displayed in Fig. 4b. Although the ordinary Hall slope has opposite signs (not shown), the sign of the AHE remains the same on both sides of the Dirac point, which is consistent with the doping dependence discussed earlier. Interestingly, in samples with widely different doping levels, not only is the relative Fermi level position with respect to the Dirac point, but also the band structure is different.[21-23] Here in one sample, the electrostatic gating only shifts the Fermi level in a fixed band structure. Therefore, the fact that the AHE sign remains unchanged across the Dirac point is robust.

Unlike in Cr- or V-doped TI, the ferromagnetism is only induced at the bottom surface of the TI layer in TI/YIG heterostructures. Nevertheless, we have demonstrated an alternative route of introducing stronger exchange interaction to TI surface states by proximity coupling. Stronger exchange should lead to a larger topological gap, therefore a higher temperature at which QAHE occurs. In the meantime, by independently optimizing the electronic properties of the TI, we can reduce disorder, especially magnetic disorder, and simultaneously tune the Fermi level position in TI without affecting the induced ferromagnetism.

## SUPPORTING INFORMATION AVAILABLE

Material growth, device fabrication and additional figures are given in the Supporting Information document. This material is available free of charge via the Internet at http://pubs.acs.org.

## AUTHOR INFORMATION

**Corresponding Authors:**

*email: jing.shi@ucr.edu

**Note:**

The authors declare no competing financial interest.

## ACKNOWLEDGMENTS

We would like to thank V. Aji for fruitful discussions, and W. Beyermann, C.N. Lau, D. Humphrey, R. Zheng, M. Aldosary & K. Myhro for the technical assistance. Work by Z.L.J. was supported by the UC Lab Fees under Award # 12-LR-237789. Work by C.T. was supported by NSF ECCS under Award # 1202559. Work by J.S. was supported by the U.S. Department of Energy (DOE), Office of Science, Basic Energy Sciences (BES) under Award # DE-FG02-07ER46351.  C. Z. C, P. W. and J. S. M. would like to thank support from the STC Center for Integrated Quantum Materials under NSF Grant No. DMR-1231319, NSF DMR Grants No. 1207469 and ONR Grant No. N00014-13-1-0301.

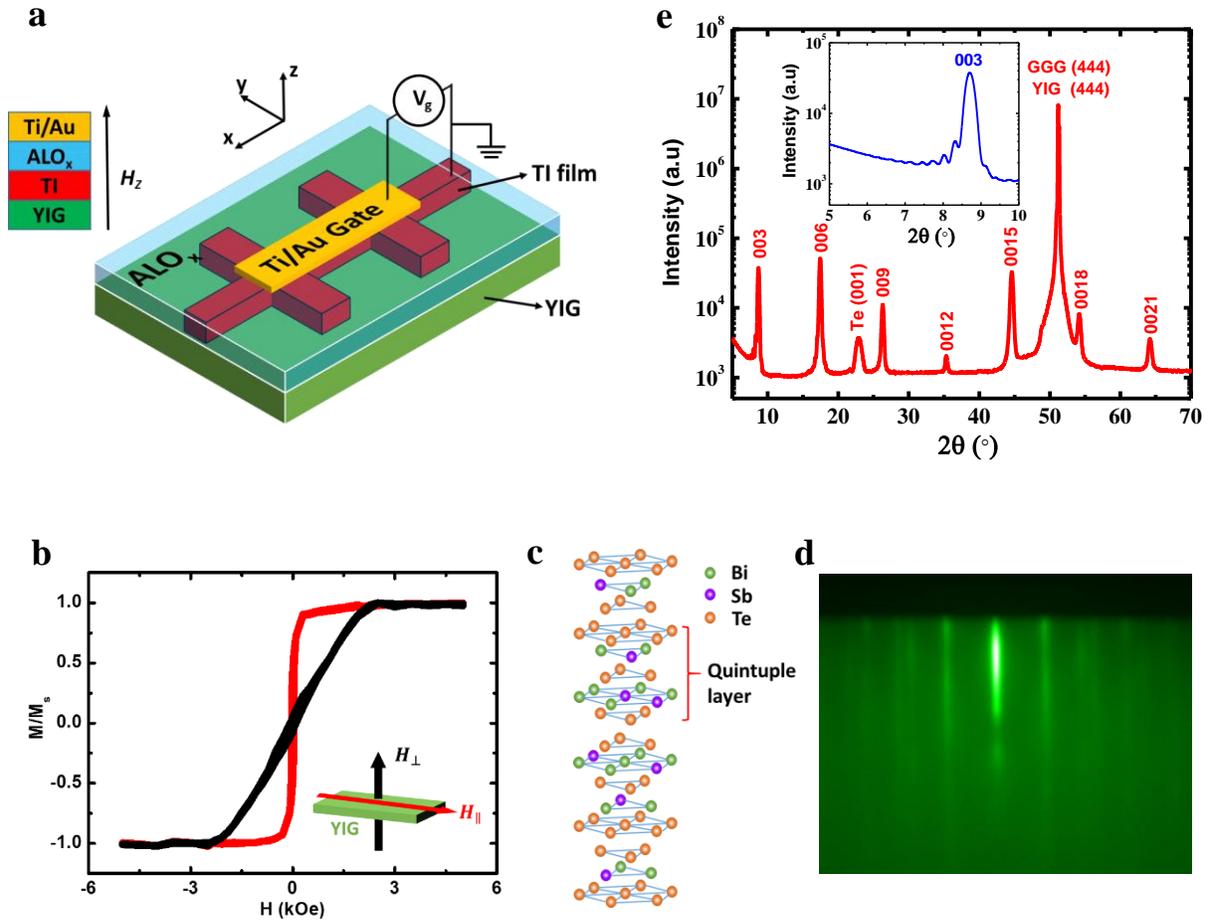

**Figure 1. Device schematics and properties of $(Bi_xSb_{1-x})_2Te_3$/YIG thin films. a,** Schematic picture of a top-gated 5QL-$(Bi_xSb_{1-x})_2Te_3$/YIG device for magneto-transport measurements with a side view. **b,** 300 K magnetic hysteresis loops measured by VSM with in-plane and perpendicular magnetic fields. The out-of-plane curve indicates a saturation field ~2700 Oe which slightly varies in different YIG samples. **c,** Tetradymite-type crystal structure of $(Bi_xSb_{1-x})_2Te_3$ consisting of quintuple layers. Bi atoms are partially replaced by Sb atoms. **d,** RHEED pattern of MBE-grown 5 QL-$(Bi_xSb_{1-x})_2Te_3$ on YIG/GGG showing highly ordered flat crystalline surface. **e,** X-ray diffraction result of a typical 20 QL-$(Bi_xSb_{1-x})_2Te_3$ grown on YIG/GGG. The inset shows a zoom-in view of the (003) peak and clear Kiessig fringes.

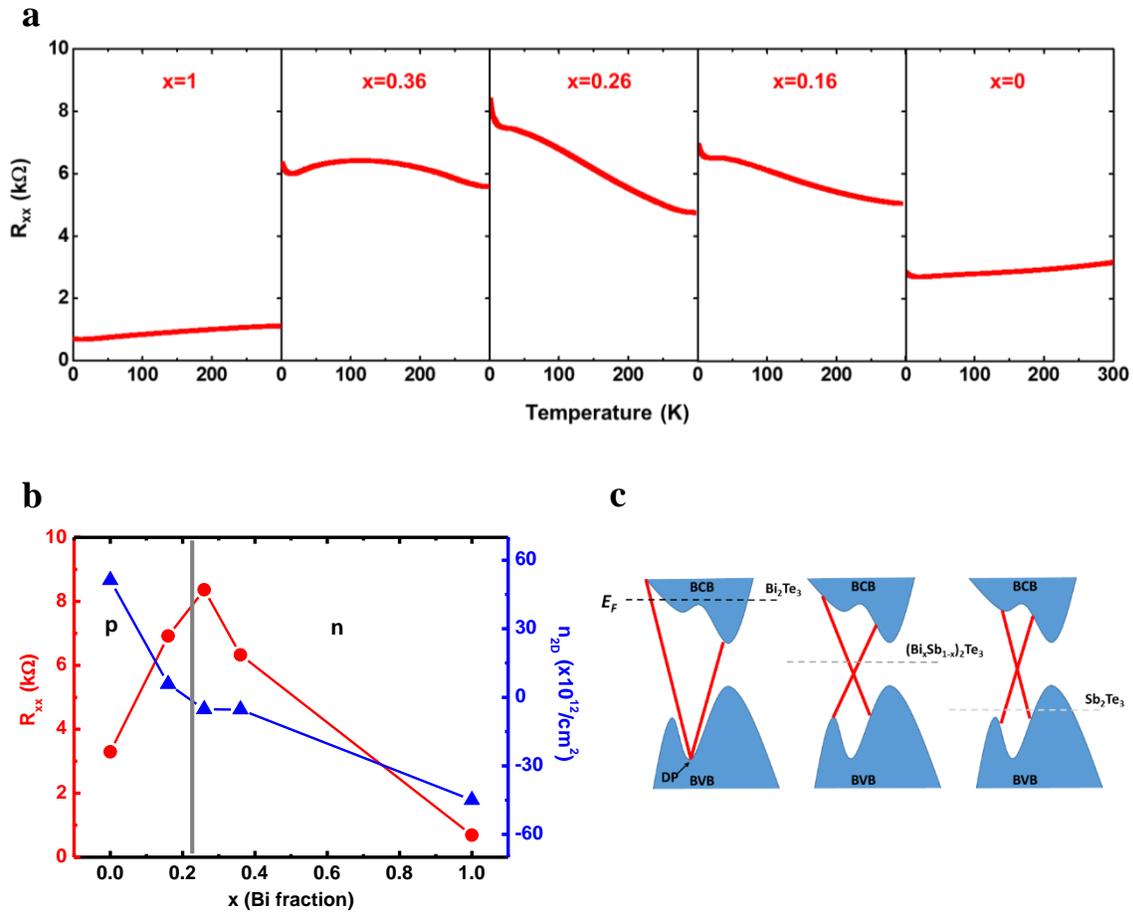

**Figure 2. Evolution of the longitudinal resistivity and carrier density with Bi fraction in five 5QL (Bi$_x$Sb$_{1-x}$)$_2$Te$_3$/YIG films. a,** Temperature dependent longitudinal resistance for five 5QL (Bi$_x$Sb$_{1-x}$)$_2$Te$_3$/YIG samples with *x* varying from 0 to 1. **b,** Longitudinal resistance and carrier density vs. Bi fraction. **c,** Schematic electronic band structure of (Bi$_x$Sb$_{1-x}$)$_2$Te$_3$ indicating the shift of the Fermi energy as *x* is varied.

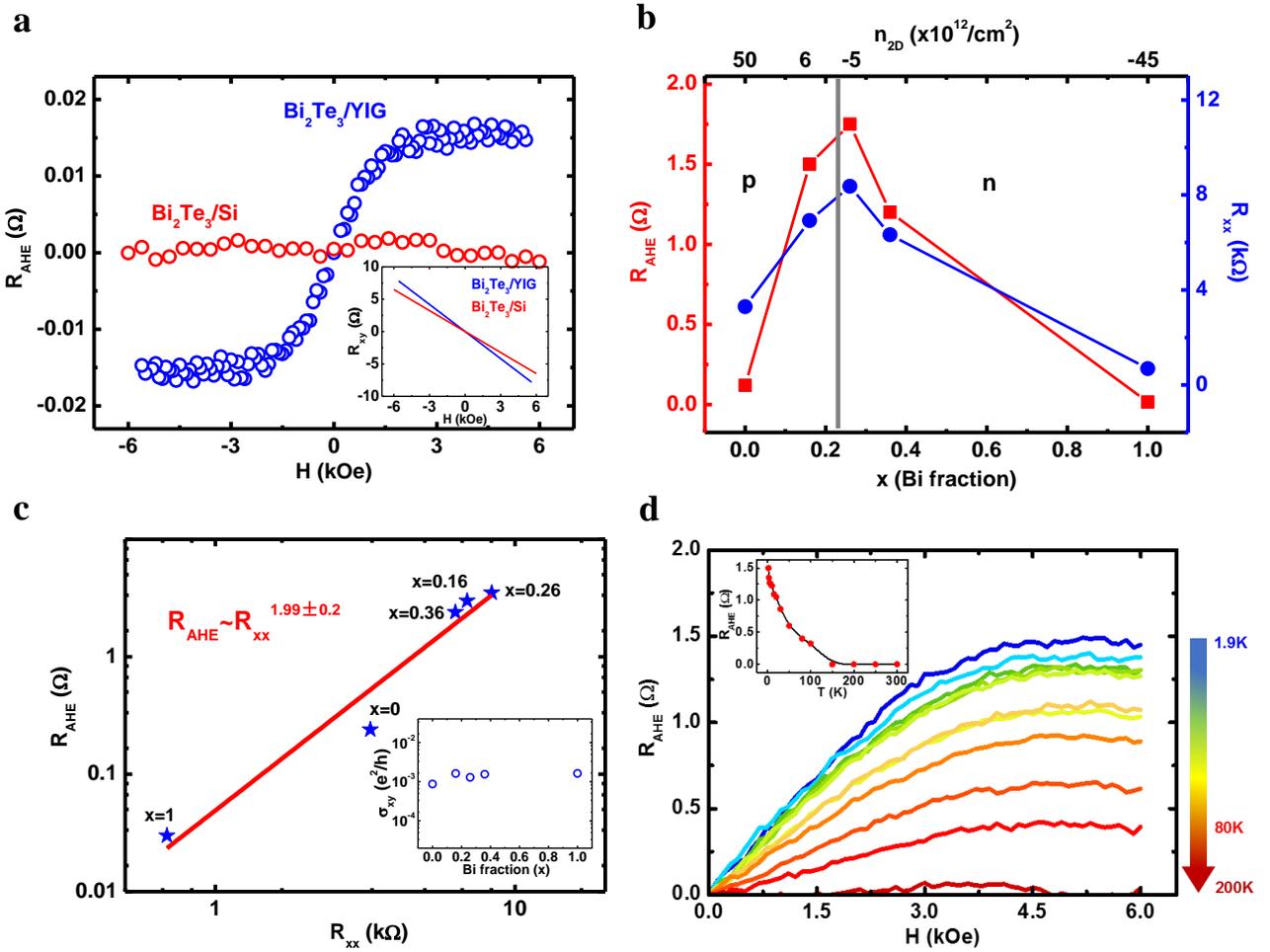

**Figure 3. Proximity induced ferromagnetism in (Bi$_x$Sb$_{1-x}$)$_2$Te$_3$/YIG films. a,** A comparison of nonlinear Hall resistivity after the linear Hall background is removed in Bi$_2$Te$_3$/YIG and Bi$_2$Te$_3$/Si. The inset shows the total Hall data for Bi$_2$Te$_3$/YIG and Bi$_2$Te$_3$/Si. **b,** AHE resistance and longitudinal resistance vs. Bi fraction (bottom axis) and carrier density (top axis). **c,** Log-log plot for AHE resistance vs. longitudinal resistance for five samples. The slope of the red line is 1.99±0.2. The inset shows σ$_{xy}$, the AHE conductivity, as the Bi fraction is varied. **d,** AHE resistance of (Bi$_{0.16}$Sb$_{0.84}$)$_2$Te$_3$/YIG sample measured from 2 to 200 K. The inset is the saturated AHE resistance as a function of the temperature showing a $T_C$ of ~150 K.

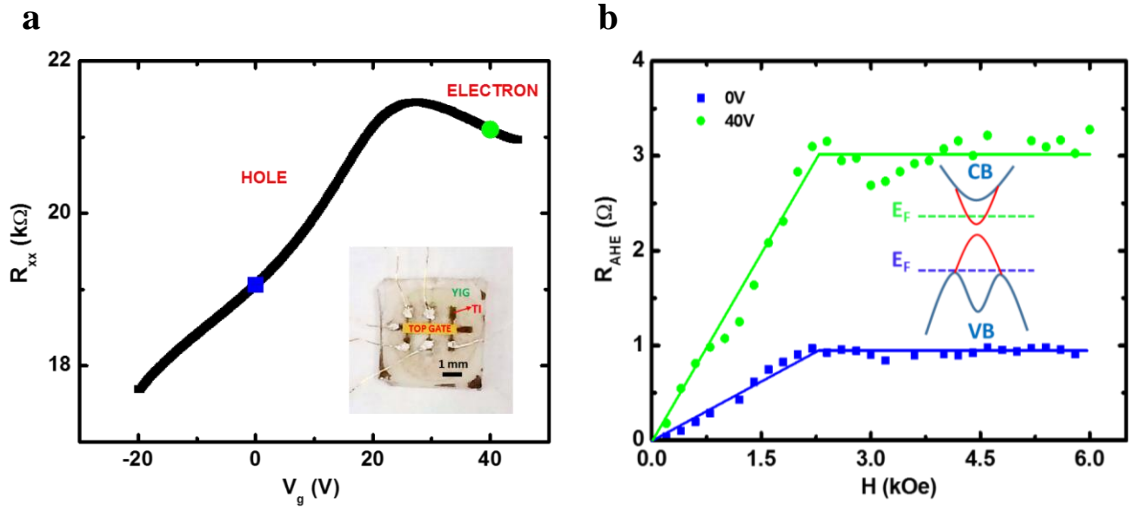

**Figure 4. AHE of $(Bi_{0.24}Sb_{0.76})_2Te_3$/YIG films tuned with gate voltage. a,** The gate voltage dependence of longitudinal resistance for $x$=0.24 sample. The blue squares and green circles represent 0 and 40 V top gate voltages, respectively. The inset is the optical image of the top-gated device. **b,** AHE resistance at different gate voltages with different carrier types: 0 V (hole side) and 40 V (electron side). Solid lines are guides for the eyes. The inset shows schematic picture of the relative Fermi level position at 0 and 40 V top gate voltages, respectively.